\documentclass[modern]{aastex63}

\usepackage{amsmath}
\usepackage{amssymb}
\usepackage{rotating}

\received{January 1, 2018}
\revised{January 1, 2018}
\accepted{January 1, 2018}

\shorttitle{Identifying microlensing events using neural networks}
\shortauthors{P. Mr\'oz}

\begin{document}

\title{Identifying microlensing events using neural networks}

\correspondingauthor{Przemek Mr\'oz}
\email{pmroz@astro.caltech.edu}

\author[0000-0001-7016-1692]{Przemek Mr\'oz}
\affil{Division of Physics, Mathematics, and Astronomy, California Institute of Technology, Pasadena, CA 91125, USA}

\begin{abstract}
Current gravitational microlensing surveys are observing hundreds of millions of stars in the Galactic bulge -- which makes finding rare microlensing events a challenging tasks. In almost all previous works, microlensing events have been detected either by applying very strict selection cuts or manually inspecting tens of thousands of light curves.
However, the number of microlensing events expected in the future space-based microlensing experiments forces us to consider fully-automated approaches. They are especially important for selecting binary-lens events that often exhibit complex light curve morphologies and are otherwise difficult to find. There are no dedicated selection algorithms for binary-lens events in the literature, which hampers their statistical studies.
Here, we present two simple neural-network-based classifiers for detecting single and binary microlensing events. We demonstrate their robustness using OGLE-III and OGLE-IV data sets and show they perform well on microlensing events detected in data from the Zwicky Transient Facility (ZTF). Classifiers are able to correctly recognize $\sim 98\%$ of single-lens events and $80-85\%$ of binary-lens events.
\end{abstract}

\keywords{Gravitational microlensing (672), Classification (1907), Neural networks (1933) }

\section{Introduction} \label{sec:intro}

Arguably, the most important legacy value of photometric surveys for gravitational microlensing events is opening-up a new branch of observational astronomy, now called \textit{time-domain astronomy}. \citet{paczynski1986} proposed to monitor the brightness of millions stars in the Magellanic Clouds over a time scale from hours to years to search for gravitational microlensing caused by hypothetical dark, compact objects in the Milky Way halo, which---as suspected at that time---may have constituted dark matter. The practical aspect of this program seemed ``formidable'' to \citet{paczynski1986} -- his proposal required frequent, repeatable observations of large high-stellar-density areas of the sky.

Setting aside practical (but essential) aspects of Paczy\'nski's proposal (photometric reductions of thousands images, measuring the brightness of millions of stars, creating efficient photometric databases, etc.), finding microlensing events in the data must have seemed challenging -- the first microlensing event detected by the Optical Gravitational Lensing Experiment \citep[OGLE;][]{udalski1993} was found among 1.1~million light curves of stars. All light curves passed through a variety of automated filters but in the last step human experts must have visually vetted hundreds of candidate objects.

The situation has not changed much since these first discoveries, despite the fact that the number of known microlensing events exceeded several thousands. Modern major microlensing surveys -- OGLE \citep{udalski2015}, MOA \citep[Microlensing Observations in Astrophysics;][]{bond2001}, and KMTNet \citep[Korea Microlensing Telescope Network;][]{kim2016} are each observing hundreds of millions of stars and selecting rare microlensing events may require laborious efforts. For example, when building their event-finding algorithm, \citet{kim2018} manually reviewed almost 400,000 light curves. The main disadvantage of such an approach -- besides the number of work hours needed to manually vet the light curves -- is the difficulty in quantifying the selection biases and measuring the event detection efficiency. Some events may be inadvertently rejected by a scanner and quantifying that effect is virtually impossible.

Broadly speaking, microlensing events may be detected either in real-time (as the event progresses) or in an offline search of archival data. The former problem being much more challenging than the latter -- due to limited amount of information in the light curves. 
In this paper, we focus on the search for microlensing events in the archival data. From the historical point of view, finding microlensing events in real-time was instrumental for the success of the field \citep[e.g.,][]{udalski1994b,gaudi2012}. The first generation surveys could not observe the sky with a high enough cadence to accurately cover light curves of interesting events, especially those by planetary systems. Thus, survey groups searched for probable microlensing events in real-time, distributed alerts, and only the most promising events were monitored with a higher cadence by follow-up teams, as advocated by \citet{gould_loeb1992}.

The observing capabilities of the current second-generation surveys have improved since then, so the role of follow-up observations -- and alerts -- has become less important. The survey telescopes monitor the central regions of the Galactic bulge with a 15--20 min cadence, which is sufficient for them to detect and characterize short-duration events and anomalies without the need of follow-up observations \citep[e.g.,][]{sumi2011,yee2012,yossi2016pl,mroz2017}. This will also be true for the planned microlensing survey with the \textit{Nancy Grace Roman Space Telescope}, which is expected to detect nearly 30,000 events \citep{penny2019}. \textit{Roman} will observe its microlensing field continuously with a cadence of 15~min. Moreover, many events detected by \textit{Roman} will be inaccessible to ground-based telescopes owing to extremely high extinction, rendering their follow-up difficult. Thus, it is vital to devise robust automated algorithms for detecting microlensing events in the survey data.

The problem of finding single-lens events in the archival data has been discussed by the number of authors \citep[e.g.,][]{alcock2000,popowski2001,sumi2003,hamadache2006,tisserand2007,wyrzyk4,sumi2011,wyrzykowski2015,mroz2017,kim2018}. Usually, they use a series of selection cuts which narrow down a sample of potential events. This introduces a problem of a balance between the size and purity of the sample -- too strict selection cuts may reject a significant fraction of genuine microlensing events, whereas the sample may be contaminated by non-microlensing light curves if the criteria are too loose.

While the problem of finding single-lens events is well understood, detecting binary microlensing events is much more challenging owing to a variety of possible light curve shapes \citep{liebig2015}. Light curves of single-lens events can be described by a simple analytical formula \citep{paczynski1986}, which renders their modeling easy and fast. On the other hand, modeling light curves of binary microlensing events requires a large amount of computational time: there are at least six highly non-linear parameters and finding the best-fitting solution requires a grid search over some of them. Thus, fitting the binary-lens models to all candidate light curves is (and likely will be) unfeasible. Any statistical studies of binary microlensing events are hampered by the lack of automated and robust detection algorithms. To our knowledge, no dedicated selection algorithms for binary-lens events are available in the literature.

The number of microlensing events expected in the future microlensing experiments forces us to consider fully-automated approaches. Here, we present two machine-learning neural-network-based classifiers for detecting single and binary microlensing events. Both algorithms are trained on a sample of microlensing events from the OGLE-IV survey \citep{mroz2017,mroz2019b}. We demonstrate their good performance on the OGLE-III sample of microlensing events. Moreover, we also demonstrate that these classifiers work well on a sample of microlensing events detected by the Zwicky Transient Facility \citep[ZTF;][]{mroz2020}, showing that they may be adapted to other experiments.

\section{Machine-learning techniques in microlensing}

Although most studies on microlensing employ a traditional approach for finding microlensing events (in which a series of selection cuts is applied to the entire sample of light curves), a few notable studies used machine-learning techniques. We would like to review them in more detail. \citet{wyrzykowski2015} used a Random Forest classifier to identify single-lens microlensing events in the OGLE-III data set. Their selection method was twofold. First, they used a series of selection cuts to narrow down a sample of possible microlensing events to $\sim 50,000$ light curves. Subsequently, they calculated 27 features for each light curve, which they used in their Random Forest classifier. The performance of the classifier was moderate: 96.7\% of objects were correctly classified, but the false-positive rate was relatively high (27.7\%). For that reason they added a second stage of classification, which allowed them to lower the false-positive rate to 6.7\%. Similarly, \citet{wyrzykowski2016} used the Random Forest classifier to select single-lens microlensing events exhibiting annual parallax effect.

Machine-learning-based techniques are also used to select a sample of microlensing events in data from the United Kingdom Infrared Telescope (UKIRT) survey \citep{chu2019}. \citet{chu2019} used three different classification methods: Random Forest, Support Vector Machine, and K-nearest Neighbor (of which the Random Forest classifier performs best, G. Bryden, priv. comm.) They extracted 66 features for every light curve, but in practice only 49 features were used for the best classifier performance. Their microlensing classifier has a 93.8\% precision and 97.8\% recall. Both \citet{wyrzykowski2015} and \citet{chu2019} trained their classifiers using a subset of manually-labeled data.

A Random Forest-based algorithm for detecting microlensing events in real-time was developed by \citet{godines2019}. They simulated light curves of constant stars, RR Lyrae and Cepheid variables, cataclysmic variable stars, and microlensing events, which they used as the training set. They calculated 47 light curve statistics for every object and then run the principal component analysis to select 44 most useful features. They achieved a 94\% accuracy on simulated light curves. \citet{godines2019} implemented their algorithm to search for ongoing microlensing events in the ZTF alert stream but its performance was affected by the lack of baseline data. 

\section{Classifiers}
\label{sec:classifiers}

\citet{mroz2017} and \citet{mroz2019b} presented two large samples of microlensing events identified in data from the OGLE-IV survey \citep{udalski2015}. Both data sets were constructed in a traditional fashion -- a series of selection cuts was applied to light curves of all objects observed by OGLE in the Galactic bulge. We were curious how efficiently this task can be performed by machine-learning techniques. Can they outperform traditional methods? First, strict selection cuts of \citet{mroz2017,mroz2019b} removed some number of genuine microlensing events from the sample. Second, the proposed criteria were designed to select point-lens point-source (PSPL) events and thus binary and anomalous events were rejected. 

\begin{figure}
\includegraphics[width=\textwidth]{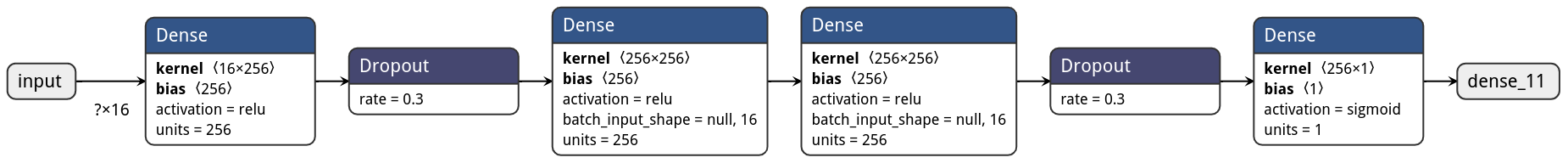}\\
\includegraphics[width=\textwidth]{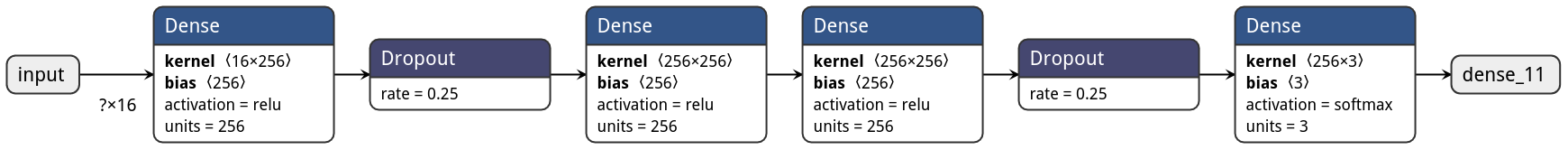}
\caption{Neural network architectures of the PSPL classifier (upper panel) and the binary lens classifier (lower panel).}
\label{fig:dnn}
\end{figure}

In this work, we use Deep Learning (DL) techniques, which are currently commonly used, both in scientific and industrial applications. For an overview of DL, we refer the reader to \citet{geron2019}. In short, our networks are composed of several layers (see Figure~\ref{fig:dnn}), every layer is composed of many neurons (``units'') and is fully connected to the next layer. Every neuron performs a simple non-linear transformation of the input data (either input features or output from an earlier layer). The network is ``trained'' by adjusting the values of its parameters (weights and biases) so that the loss function (quantifying the difference between the network's predictions and input labels) is minimized. We additionally use standard ``dropout layers'', which prevent the network from overfitting.

We did not attempt to classify all OGLE light curves (including those of constant or periodic variable stars). We used the methods of \citet{mroz2017} and \citet{mroz2019b} to preselect a sample of light curves ``enriched'' in microlensing events. We chose objects showing at least three consecutive data points (``bump'') that are at least $3\sigma_{\rm base}$ above the baseline flux $F_{\rm base}$, where $F_{\rm base}$ and $\sigma_{\rm base}$ are calculated using data points outside a 720 day window centered on the event (after removing $5\sigma$ outliers). We imposed the following four conditions: i) $\chi^2_{\rm out}/\mathrm{dof} \leq 2$, ii) $\chi_{3+} \geq 32$, iii) $n_{\rm DIA} \geq 3$, and iv) $A \geq 0.1$ mag. Here, $\chi^2_{\rm out}=\sum_i (F_i-F_{\rm base})^2/\sigma_i^2$ for data points outside the window (it quantifies variability in the baseline), $\chi_{3+}=\sum_i (F_i-F_{\rm base})/\sigma_i$ for data points within a bump, $n_{\rm DIA}$ is the number of data points within a bump detected on the subtracted images, and $A$ is the amplitude of the event. We additionally removed all candidates that were located close to each other and magnified in the same images -- these are spurious detections. See \citet{mroz2017,mroz2019b} for a more detailed description of the method and selection cuts. Our previous tests indicate that the vast majority of genuine microlensing events ($\sim 95\%$) passes these criteria \citep{mroz2019b}. 

We trained our DL algorithms on that sample ``enriched'' in microlensing events. We constructed two classifiers: the PSPL classifier is supposed to distinguish PSPL events from all other objects, whereas the binary-lens classifier is designed to recognize both PSPL and binary (anomalous) microlensing events.

\subsection{Features and training sample}
\label{sec:features}

We calculated 16 features for each light curve\footnote{The code for calculating light curve features is available at \url{https://github.com/przemekmroz/ml_microlensing}.}. We chose to use light curve features connected to a microlensing model rather than generic features, such as those used in previous works \citep[e.g.,][]{godines2019}. We also opted to choose features that do not explicitly depend on the cadence of observations.
Five features were calculated as part of the event preselection, these are: the mean brightness in the baseline, $\chi^2_{\rm out}/\mathrm{dof}$, amplitude and duration of the event, and presence of additional bumps in the light curve. Remaining features are related to the best-fitting PSPL model: $\log t_{\rm E}$ (logarithm of Einstein timescale), $u_0$ (impact parameter), $f_{\rm s}$ (dimensionless blending parameter), flag indicating whether the fit converged, and seven values $\chi^2/N$ calculated for all data points and in the time ranges $|t-t_0| < t_{\rm E}$, $|t-t_0| < 2 t_{\rm E}$, $t_0-t_{\rm E}<t<t_0$, $t_0<t<t_0+t_{\rm E}$, $t_0-2t_{\rm E}<t<t_0$, and $t_0<t<t_0+2t_{\rm E}$ (where $t_0$ is the time of the closest approach between the lens and source, and $N$ is the number of data points in a given range).

The training sample consists of 32,378 objects that were selected as possible microlensing events by \citet{mroz2017} and \citet{mroz2019b}. Our sample is very representative -- it includes microlensing events from high-cadence OGLE-IV fields from the seasons 2010--2015 and events from low-cadence fields from 2010--2018. All light curves were manually classified by a human expert into three categories: 1) JUNK (non-microlensing light curves: artifacts, cataclysmic variables, flaring stars, etc.), 2) PSPL (point-source point-lens events), 3) BINARY (binary and anomalous microlensing events). The apparent PSPL events with unphysical model parameters ($t_{\rm E} > 500$\,d and $f_{\rm s} \approx 0$, indicating large uncertainties and strong correlations between the measured parameters) or noisy light curves ($\chi^2/N>2$ for the entire light curve) were included in the category ``JUNK'' -- such events are not useful in the scientific analyses.
When we run the first versions of the classifiers, we visually inspected misclassified light curves. In many cases, the initial labels turned out to be incorrect due to a scanner mistake -- we thus iteratively amended the labels. The final training sample contains 19,867 light curves labeled as ``JUNK'', 11,701 -- ``PSPL'', and 810 -- ``BINARY''. The training sample was divided into training, validation, and test sets, which include 80\%, 10\%, and 10\% of all objects, respectively.

\subsection{Training process}

The classifiers\footnote{The trained weights for the classifiers are available in \textsc{HDF5} format at \url{https://github.com/przemekmroz/ml_microlensing}.} were implemented using \textsc{TensorFlow} software and its \textsc{Keras API} \citep{abadi2016,chollet2015}. We used the binary cross-entropy and sparse categorical crossentropy loss functions (for the PSPL and binary-lens classifiers, respectively), a batch size of 32, and the Adam optimizer \citep{kingma2014}. The training process was stopped if no improvement in validation accuracy was observed. We used the \textsc{keras-tuner} library \citep{omalley2019} to find the optimal values of network hyperparameters: number of layers and neurons, dropout rates, activation functions, and initial learning rate. The final neural net architectures are presented in Figure~\ref{fig:dnn}. Both architectures are very similar: they differ by dropout rates and the activation function for the output layer. The PSPL classifier returns a score from 0.0 to 1.0, single-lens events are more likely to have their score near 1. The binary-lens classifier returns three numbers from 0.0 to 1.0, which represent probabilities of three possible outcomes (JUNK, PSPL, or BINARY).

The number of binary events in the training sample was significantly smaller than the number of other objects. We thus resampled the training set by oversampling the minority class (so the number of the resampled binary events was equal to the number of PSPL events in the training set). This ensures that binary events are included in each training batch which makes it easier to train the model. 

\begin{table}
\centering
\begin{tabular}{lccc}
\hline
dataset & accuracy & precision & recall \\
\hline
validation set & 0.980 & 0.962 & 0.981 \\
test set & 0.977 & 0.962 & 0.975 \\ \hline
\end{tabular}
\caption{Performance of the PSPL classifier on the validation and test sets.}
\label{tab:pspl}
\end{table}

\subsection{Classifiers performance}

We split objects detected by \citet{mroz2017,mroz2019b} into training, validation, and test sets. The training and validation sets were explicitly used in the training process, so the test set provides us the most reliable measures of the classifiers' performance (classifiers did not ``see'' light curves from the test set). 

The PSPL classifier has a superb completeness and purity (Table \ref{tab:pspl}). The accuracy (fraction of correctly classified objects) of the classifier is 0.980 and 0.977 for the validation and test sets, respectively. Similarly, its recall (fraction of correctly classified PSPL events) is high -- 0.981 and 0.975 for the validation and test sets, respectively. The precision of the classifier (which quantifies the purity of the sample) is slightly lower (0.962). 

\begin{figure}
\centering
\includegraphics[width=.7\textwidth]{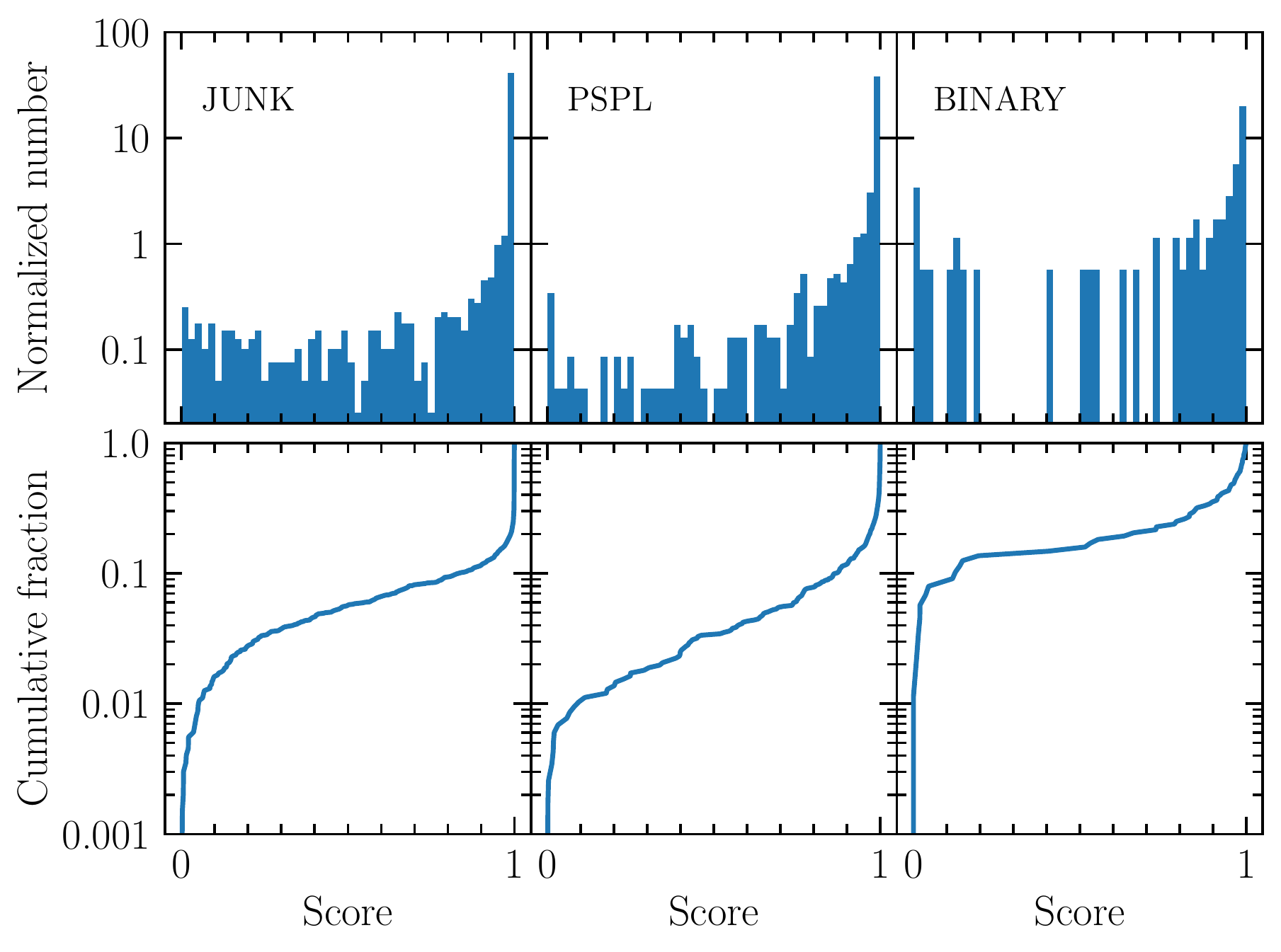}
\caption{Performance of the binary lens classifier on the test set. Each column presents the score distribution for objects classified as JUNK, PSPL, or BINARY by an expert.}
\label{fig:binary1}
\end{figure}

\begin{figure}
\centering
\includegraphics[width=.5\textwidth]{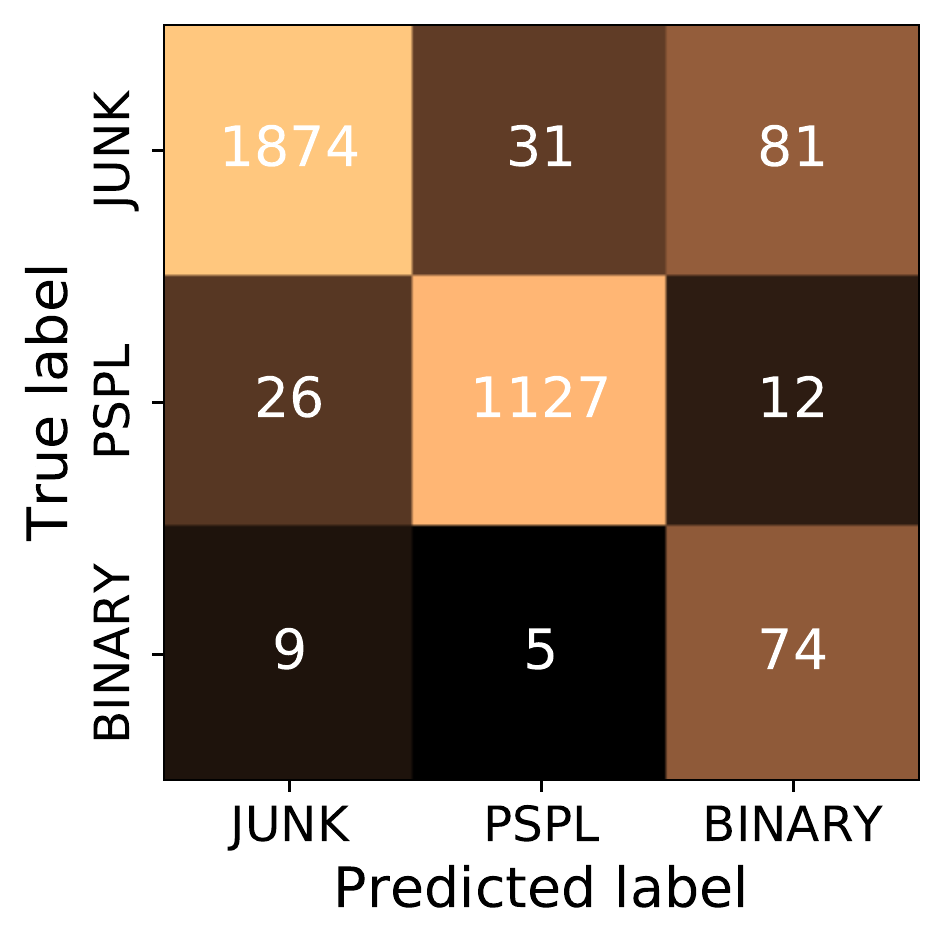}
\caption{Performance of the binary lens classifier on the test set.}
\label{fig:binary2}
\end{figure}

The performance of the binary-lens classifier is slightly poorer (Figures \ref{fig:binary1} and \ref{fig:binary2}). 96.7\% (84.1\%) of PSPL (binary) events from the test set were correctly classified. There is some confusion between PSPL and binary events -- 5.7\% of binary-lens events were classified as PSPL events. Although the binary-lens classifier is able to recognize the majority of binary events, it has a relatively low precision ($\sim 44\%$). A small fraction of non-microlensing light curves ($81/1986 = 4\%$) was incorrectly classified as binary events. Because binary-lens events are rare, this leads to a large contamination. However, the main value of the classifier is that the sample of possible binary events in the test set was reduced from 3239 to 167, that is, by $95\%$, which aids the manual classification. We envision that our classifier may be used in statistical studies to vet light curves of simulated events -- its primary role is to recognize binary events, not to remove non-microlensing light curves.

Poleski et al. (in prep.) in their search for wide-orbit planets have recently manually classified light curves of a subset of possible microlensing events detected in the OGLE-III data by the event-finding algorithms of \citet{mroz2017,mroz2019b}. They marked 1701 objects as likely PSPL events (``events which seem useful for planet detection efficiency calculations'') and 92 light curves as binary-lens events (``obvious binary-lens or binary-source event, but not wide-orbit planet''). For 1663/1701 events (that is, 97.8\%), the PSPL classifier score is larger than 0.5. Similarly, the binary-lens classifier correctly recognized 1667/1701 (98.0\%) PSPL events and 73/92 (79.3\%) binary-lens events.

One may argue that our classifiers work well because they were trained on the OGLE data and so it is not surprising that their performance on the OGLE data is excellent. As an independent test, we checked the classifiers' performance on microlensing events detected in the first year of ZTF observations \citep{mroz2020}. That sample includes 23 PSPL events, one PSPL event with strongly pronounced annual microlens parallax effect (ZTF19aainwvb), and six binary events. The PSPL classifier correctly classified 22/23 single-lens events (95.7\%), one event (ZTF18abqbeqv) has a low score (0.159) because the best-fitting model converged to implausibly long $t_{\rm E} > 1000$\,d (parameters of that event are poorly constrained and strongly correlated, see \citealt{mroz2020}). The binary-lens classifier correctly recognized 22/23 PSPL events (with the exception of ZTF18abqbeqv) and 5/6 binary events. ZTF18abqazwf was not recognized as a binary-lens event but this object has an atypical double-peak light curve and is located relatively far from the Galactic plane, so it may be some unusual variable star (\citealt{mroz2020} classified it as a ``possible'' microlensing event). We conclude that our classifiers perform well on ZTF microlensing events.

\begin{figure}
\centering
\includegraphics[width=.49\textwidth]{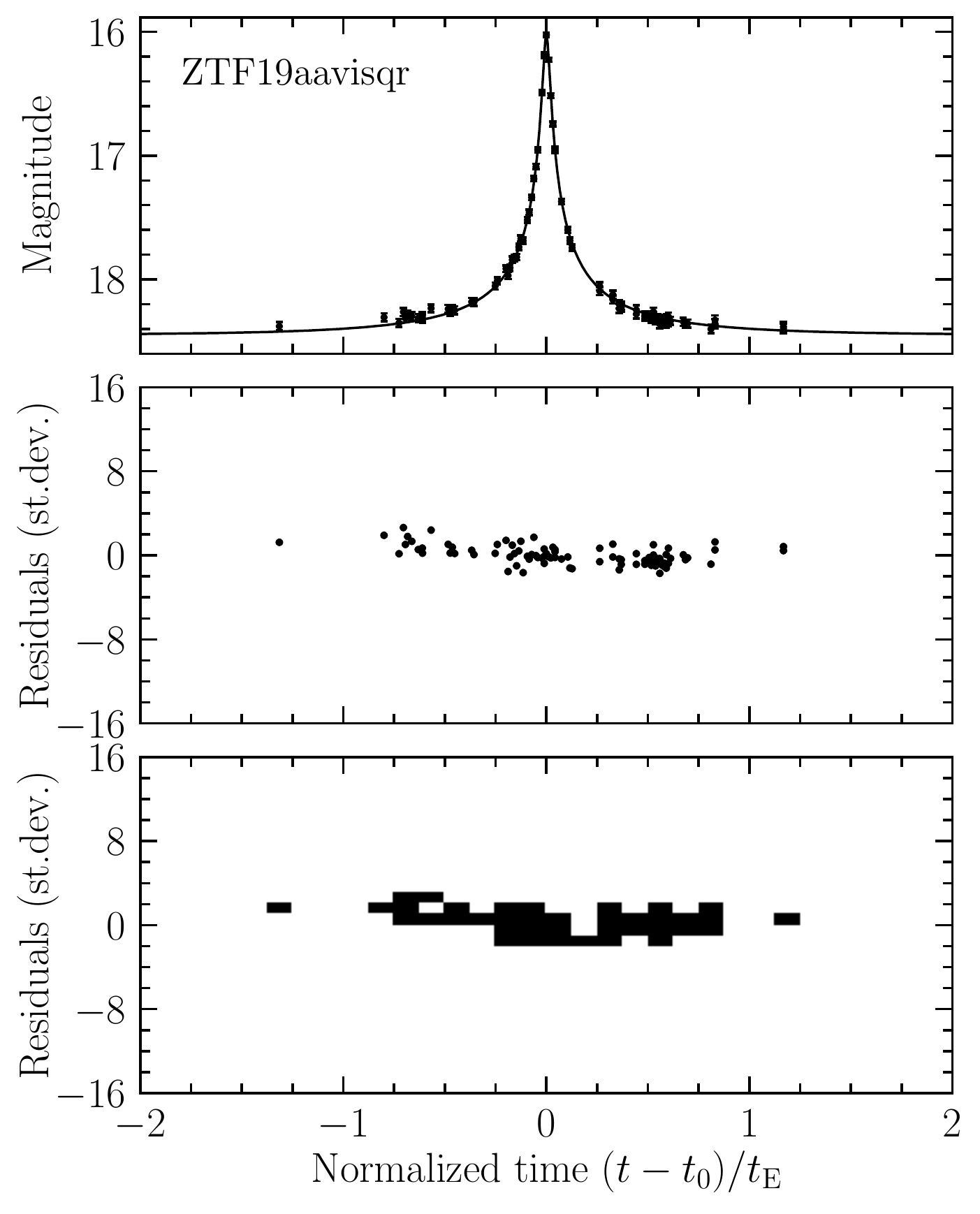}
\includegraphics[width=.49\textwidth]{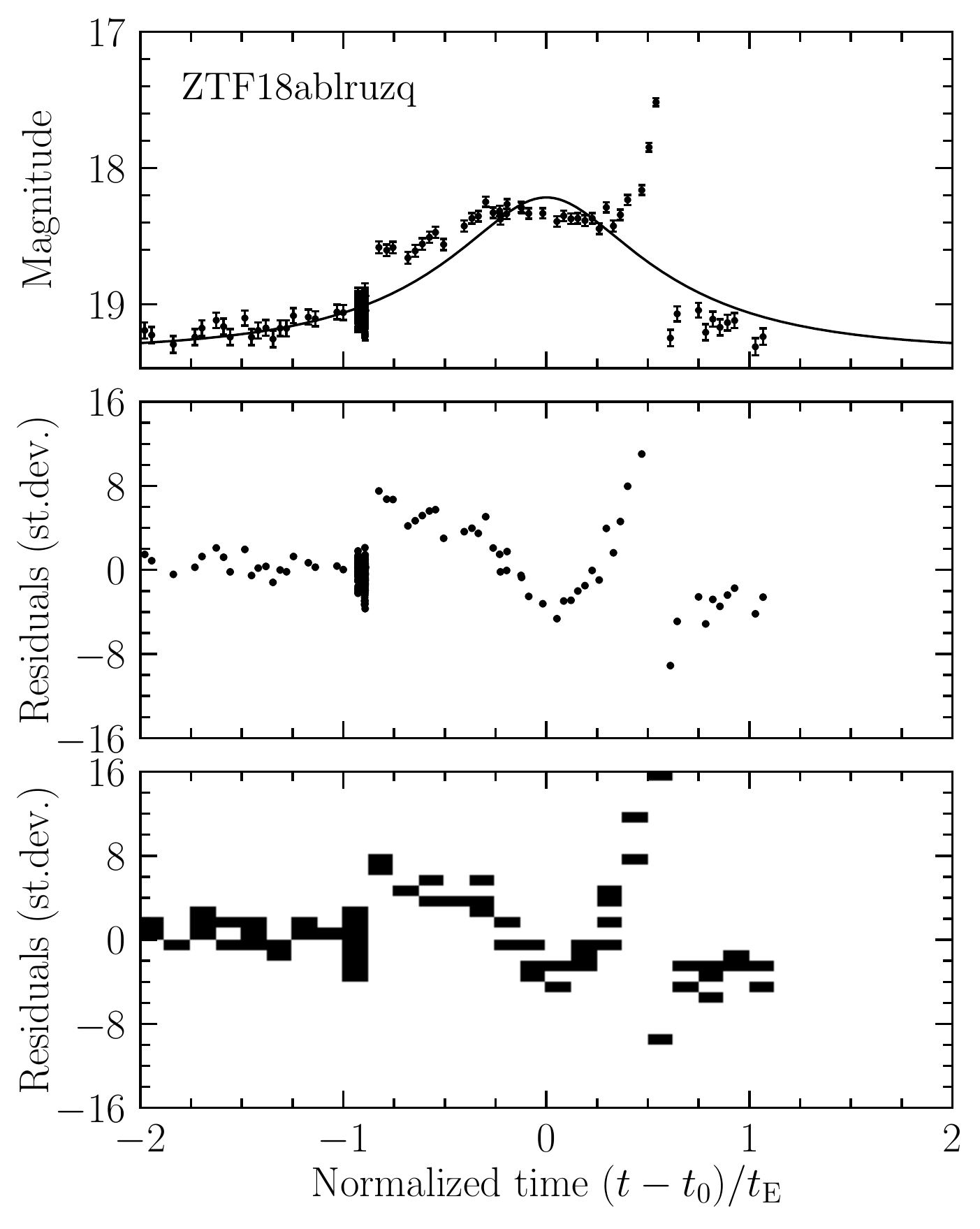}
\caption{How to turn a microlensing light curve into a 2D image? Microlensing events may last from hours to hundreds of days but the time axis can be normalized to the Einstein timescale of the event. On the $y$ axis (middle and lower panels), we plot residuals (expressed in standard deviations) from the best-fitting PSPL model (black line in upper panels). If the light curve is a genuine PSPL event, nearly all data points will be in the range $-3<y<3$. We then discretize the middle plots into $32 \times 32$ bins (lower panels), which we feed into a CNN classifier. Left column presents a PSPL event, right -- a binary-lens event.}
\label{fig:cnn}
\end{figure}

\subsection{CNN classifier}

We also explored the possibility of using convolutional neural network (CNN) classifiers to recognize microlensing events in the data. CNNs are commonly used to analyze images, we thus needed to convert a microlensing light curve into a 2D representation. This is a challenging task because microlensing events may last from hours to hundreds of days and have amplitudes from $\sim 0.1$ to $>5$ mag. We devised the following method (see Figure~\ref{fig:cnn} for an illustration of the method).

We first fitted the microlensing PSPL model to the light curve and calculated residuals from the best-fitting model. Each microlensing event has a characteristic timescale (the Einstein timescale $t_{\rm E}$), which we used to normalize the time axis. To construct the 2D representation of the light curve, we used data points collected in the time range $-2 \leq (t-t_0)/t_{\rm E} \leq 2$ so they cover almost the entire magnified part, regardless of the event duration (Figure~\ref{fig:cnn}). Subsequently, we normalized the residuals from the best-fitting PSPL model by their uncertainties ($y_i = r_i/\sigma_i$, where $r_i$ is the flux difference between the data and the model and $\sigma_i$ is the uncertainty of the $i$th data point). If $|y_i|>16$, we assumed $y_i=-16$ or $y_i=15$. If the light curve is a genuine PSPL event, nearly all data points are in the range $-3<y_i<3$. For binary-lens events, residuals are correlated in time (as shown in Figure~\ref{fig:cnn}). Finally, we discretize the residuals into $32\times 32$ bins. If there is at least one data point within a bin, we assign it a value of 1, 0 otherwise. The resulting 2D image representations are presented in the lower panels of Figure~\ref{fig:cnn}. Light curves of PSPL events are represented by a thin horizontal line, whereas other objects have more complicated shapes. 

We used both 2D light curve representations and 16 features (Section~\ref{sec:features}) as an input of our CNN classifier. However, despite being considerably more complex, the performance of the CNN classifier was similar (or even poorer in the case of binary events) to that of PSPL and binary-lens classifiers. For example, the CNN classifier correctly recognized 97.4\% (65.2\%) of PSPL (binary) events from the test set (the binary-lens classifier achieved a recall of 96.7\% and 84.1\%, for the same test set, respectively).

One possible issue with the CNN classifier is the difficulty of constructing a 2D representation of the light curve, especially if there are gaps in the data. The 2D representation may also depend on the cadence of observations (note that the light curve features used in our classifiers do not explicitly depend on that). Hence, using the CNN classifier for the OGLE data does not result in the significantly better performance. However, the proposed 2D representation of light curves and the CNN classification may be used for finding microlensing events in data from the \textit{Roman Space Telescope}. \textit{Roman} will observe the sky nearly continuously with a uniform cadence, thus, its data products are much better suited for the CNN classification.

\section{Discussion}

In the first paragraph of Section~\ref{sec:classifiers} we asked ourselves whether DL classifiers can outperform traditional methods of selecting microlensing events. Comparing DL results with those of \citet{mroz2017,mroz2019b} is difficult because we used their microlensing events as the training sample. Our tests, based on a subsample of events not used in the training process, indicate that both classifiers are able to correctly classify $97.5-98.0\%$ of PSPL events. The training sample included 11,701 PSPL events, from which we should be able to select $\sim 11,400$, whereas the sample of \citet{mroz2019b} contains only 8002 events (that is, DL classifiers were able to select $\sim 40\%$ more events). However, \citet{mroz2019b} applied some physically-motivated cuts (such as limits on $u_0$ and source magnitude), so the real improvement is smaller than 40\%. On the other hand, the samples of \citet{mroz2017,mroz2019b} were very pure ($\sim 99.5\%$), whereas our classifiers have a lower precision ($\sim 96\%$). Thus, the answer to the question raised in Section~\ref{sec:classifiers} depends on the scientific application of the data set. If a large sample size is needed, machine-learning techniques are better. If high purity is required, traditional methods may perform better.

The performance of our DL classifiers is as good as that of traditional techniques of finding microlensing events (and in some cases classifiers were even better than a human). This has several reasons. First, our training sample is based on real data and includes a significant fraction of artifacts and other non-microlensing light curves. Some previous classifiers were trained on artificially-created data which -- when dealing with unknown objects -- returned spurious results. Instead of creating synthetic light curves ``from the scratch'', a better strategy may involve injecting microlensing signal into real light curves of constant stars (see \citet{mroz2019b} for details of such simulations). Such approach has an advantage that it preserves original noise in the data, which may be non-Gaussian and otherwise difficult to simulate. 

Another explanation of our classifiers' performance in the quality of the training and test data. In particular, their noise properties are well understood \citep{skowron2016} which enables us to use $\chi^2$ per degree of freedom as a robust and transferable statistics. Moreover, the light curve features used in our classifiers do not explicitly depend on cadence and number of observations. Thus, the classifiers -- trained on OGLE light curves -- work well at recognizing microlensing events in the ZTF data.

The current microlensing surveys have worked well without using machine-learning techniques to identify microlensing events. However, as we demonstrated, neural-network classifiers may be used to construct unbiased large samples of single-lens events or to recognize binary-lens events, which are otherwise difficult to identify using traditional methods. Our classifiers are able to correctly recognize $\sim 98\%$ of single-lens events and $80-85\%$ of binary-lens events in the test sets. 
As we expect that the future space-based surveys should be able to detect an order of magnitude more events than the current ground-based experiments, the use of DL techniques may be necessary to achieve the maximum science returns. High-cadence continuous observations from space are especially suited for machine-learning classification.

\section*{Acknowledgements}

This work has made use of data from the OGLE survey. We would like to thank OGLE observers for their contribution to the collection of the photometric data used in this paper. We would like to thank Radek Poleski for sharing his classifications of OGLE microlensing events and Dmitry Duev for discussions on neural networks.
We thank Dmitry Duev, Radek Poleski, and Andrzej Udalski for their comments on the manuscript. 

\bibliographystyle{aasjournal}
\bibliography{pap}

\end{document}